==================

\documentstyle[12pt]{article}

\newcommand{\NP}[1]{{\em Nucl.Phys.~B} {\bf {#1}}}
\newcommand{\PL}[1]{{\em Phys. Lett.} {\bf {#1}}}

\newcommand{\Ann}[1]{{\em Ann. Phys.} {\bf {#1}}}
\newcommand{\SU}{{\rm SU}}
\newcommand{\U}{{\rm U}}
\newcommand{\eq}{\begin{equation}}
\newcommand{\eqend}{\end{equation}}
\newcommand{\eqa}{\begin{eqnarray}}
\newcommand{\nonueqa}{\begin{eqnarray*}}
\newcommand{\eqaend}{\end{eqnarray}}
\newcommand{\nonueqaend}{\end{eqnarray*}}
\newcommand{\nonu}{\nonumber \\ \nopagebreak}
\newcommand{\bma}[1]{\begin{array}{#1}}
\newcommand{\ema}{\end{array}}

\newcommand{\Ref}[1]{(\ref{#1})}
\newcommand{\dd}{\mbox{${\rm d}$}}
\newcommand{\ee}[1]{\mbox{{\rm e}}^{#1}}
\newcommand{\ii}{{\rm i}}

\newcommand{\R}{{\sf I} \! {\sf R}}

\newcommand{\Z}{{\sf Z} \! \! {\sf Z}}

\newcommand{\f}{\frac}
\renewcommand{\vec}[1]{\mbox{\boldmath ${#1}$}}

\newcommand{\ccr}[2]{{[} {#1},{#2} {]} }        
\newcommand{\car}[2]{{\{} {#1},{#2} {\}} }        

\newcommand{\tra}[1]{{\rm tr} ({#1})}          
\newcommand{\normal}[1]{:\!{#1}\!:}
\newcommand{\GG}{\underline{G}}
\renewcommand{\gg}{\underline{g}}
\newcommand{\HH}{\vec{{\rm H}}}
\renewcommand{\AA}{{\rm A}}
\newcommand{\BB}{{\rm B}}
\newcommand{\CC}{{\rm C}}
\newcommand{\DD}{{\rm D}}
\newcommand{\cL}{{\cal L}}
\newcommand{\cH}{{\cal H}}

\begin{document}
\pagestyle{empty}
\begin{flushright}
{\large UBCTP 92-019\\
July 1992}
\end{flushright}
\vspace{0.2cm}

\renewcommand{\thefootnote}{\alph{footnote}}
\begin{center}
{\Large \bf Gauge theories on a cylinder} \\
\vspace{0.4 cm}

{\large Edwin Langmann}\footnote{supported by the ``Fonds zur
F\"orderung der wissenschaftlichen Forschung'' under the contract Nr.\
J0599-PHY}$^,$\footnote{on leave of absence of Institute of
Theoretical Physics, TU-Graz, Austria} \\ {\large and Gordon W.
Semenoff}\footnote{supported in part by the Natural Sciences and
Engineering Research Council of Canada.} \\
\vspace{0.2 cm}
{\em Department of Physics, The University of British Columbia\\
Vancouver, B.C., V6T 1Z1, Canada }\\

\end{center}
\setcounter{footnote}{0}
\renewcommand{\thefootnote}{\arabic{footnote}}

\vspace{0.1cm} \noindent

\begin{abstract}
Gauge theory with massless fermions on a cylinder (= (1+1)
dimensional spacetime with compact space $S^1$) is studied in the
Hamiltonian framework. Without using a gauge fixing condition, gauge
degrees of freedom are eliminated by explicitly solving Gauss' law.
The resulting effective Hamiltonian describes interacting fermions
coupled to a finite number of quantum mechanical variables
representing the physical Yang-Mills degrees of freedom. The method
can be trivially extended to other gauge theories with matter on a
cylinder.
\end{abstract}

\newpage

\pagestyle{plain}
\setcounter{page}{1}

A rigorous construction and investigation of gauge theories with
matter in (3+1) dimensions is beyond present day's knowledge. (1+1)
dimensional models are much simpler and can be used as a testing
ground to get more insight into gauge theories on a sound mathematical
basis.

The identification of the physical degrees of freedom of non-Abelian
gauge theories is highly non-trivial and crucial for understanding the
physics of these models. Though trivial on a plane (= spacetime
$\R\times\R$), pure YM theory on a cylinder (= spacetime
$S^1\times\R$) is non-trivial and has a finite number of physical
degrees of
freedom. This is
most easily seen in the Hamiltonian framework: Here, the
temporal component $A_0$ of the YM field is not dynamical but only a
Lagrange multiplier enforcing Gauss' law (= invariance under all
static gauge transformations), and the only gauge invariant quantities
one can construct at some fixed time from the spatial component $A_1$
of the YM-field are the eigenvalues of the Wilson loop (holonomy)
$W[A_1]$ over the whole space.

In a recent interesting paper Rajeev \cite{Rajeev} presented a method
to eliminate gauge degrees of freedom in pure YM theory on a cylinder
and reduced it to a model for a free ``particle'' moving on a group
manifold $\GG$ (the structure group of the YM field) which can be solved
exactly \cite{Dowker} ($q=W[A_1]\in \GG$ can be interpreted as
position of the ``particle'' which due to invariance under rigid
gauge transformations, $q\to h^{-1}qh$ $\forall h\in \GG$, effectively
moves only on the Cartan subgroup of $\GG$ \cite{Dowker}). The
approach of \cite{Rajeev} is very elegant as no gauge fixing condition
is imposed but Gauss' law is explicitly solved resulting in an
effective Hamiltonian $\HH_{\rm eff}$ on the physical Hilbert space
$\cH_{\rm phys}$ of the model.

In this Letter we extend the method of Rajeev \cite{Rajeev} to
massless fermions coupled to a Yang-Mills field with gauge group
$\GG=\SU(N)$ or $\U(N)$ on a cylinder.  The construction is done on a
semiclassical level with the fermion field algebra in the naive
(unphysical) representation (no filled Dirac sea) as this simplifies
the arguments and allows for a straightforward extension to other
gauge theories with matter on a cylinder.  We note that a rigorous
construction of this model and the elimination of the gauge degrees of
freedom can be done also on the full quantum level, and that the final
result does not depend on whether one goes to the full quantum level
before or after this elimination (we hope to report on that in a
future publication).

We start with the Lagrangian
density\footnote{$\partial_\nu\equiv\partial/\partial x^\nu$ with
$x^0\equiv t$ (time), $x^1\equiv x\in [0,2\pi)$ (spatial coordinate);
our metric tensor is $g_{\nu\mu}=diag(1,-1)$}
\eq
\cL = \bar{\psi}\gamma^\nu(-\ii\partial_\nu +A_\nu)\psi -
\f{1}{2e^2}\tra{F_{\nu\mu}F^{\nu\mu}}
\eqend
with\footnote{the $T^a=(T^a)^*$ are the generators of the Lie algebra
$\gg$ of $\GG$ in the fundamental representation normalized to that
$\tra{T^aT^b}=\delta^{ab}/2$ ($\tra{\cdot}$ is the $N\times N$-matrix
trace)} $A_\nu\equiv A_\nu^aT^a$ the YM field, $F_{\nu\mu}\equiv
\partial_\nu A_\mu -\partial_\mu A_\nu +\ii\ccr{A_\nu}{A_\mu}$ with
$\nu,\mu\in\{0,1\}$ spacetime indices, $\psi$, $\bar{\psi}\equiv
\psi^*\gamma^0$ the fermion fields, and $e$ the coupling constant;
more explicitly, $\psi^{(*)}(x)\equiv\psi^{(*)}_{\sigma,\AA}(x)$,
$T^a\equiv T^a_{\AA\BB}$, $\gamma^\nu\equiv\gamma^\nu_{\sigma\sigma'}$
with $\sigma,\sigma'\in\{1,2\}$ and $\AA,\BB\in\{1,2\ldots N\}$ the spin
and color indices, respectively. By the usual canonical procedure
\cite{Sundermeyer} we obtain the Hamiltonian (we assume periodic
boundary conditions for all fields)
\eq
\label{a1}
\HH =\int_0^{2\pi}\dd{x}\left(\tra{e^2 \Pi_1(x)^2 - 2 A_0(x)G(x)} +
\psi^*(x)\gamma_5(-\ii\partial_1+A_1(x))\psi(x)\right)
\eqend
where $\Pi_\nu^a(x)=F_{0\nu}^a(x)/e^2$ and $\ii\psi^*_{\sigma,\AA}(x)$
are the canonical momenta to $A^{a,\nu}(x)$ and $\psi_{\sigma,\AA}(x)$,
respectively, $\gamma_5=-\gamma^0\gamma^1$,
and\footnote{we use the notation $G(x)\equiv G^a(x)T^a$
and similarly for $A_\nu$, $\Pi_\nu$, $\rho$ etc.}$^{,}$\footnote{
$\ccr{\cdot}{\cdot}$
is the commutator and $\car{\cdot}{\cdot}$ the anticommutator}
\eq
\label{a2}
G(x)\equiv -\partial_1\Pi_1(x) - \ii\normal{\ccr{A_1(x)}{\Pi_1(x)}} +
\rho(x)
\eqend
with $\rho^a(x)=\psi^*(x)T^a\psi(x)$; the normal ordering
$\normal{\cdots}$ means that terms coming from the non-commutativity
of $A_1^a(x)$ and $\Pi_1^b(y)$ have to be discarded.  Moreover, we have
the following canonical (anti-) commutator relations
\eqa
\ccr{A_\nu^a(x)}{\Pi_{\mu}^b(y)}=-\ii g_{\nu\mu}\delta^{ab}\delta(x-y)\nonu
\car{\psi_{\sigma,\AA}(x)}{\psi^*_{\sigma',\BB}(y)}
=\delta_{\sigma\sigma'}\delta_{\AA\BB}\delta(x-y)
\eqaend
with the other (anti-) commutators vanishing as usual.
{}From the primary constraint $\Pi_0(x)\simeq 0$ we get the secondary
constraint $G(x)\simeq 0$ (Gauss' law).  Following Rajeev \cite{Rajeev} we
introduce the Wilson line $S(x)\equiv S_{\AA\BB}(x)$ as the (unique)
solution of
\eq
\label{a5}
\left(\partial_1 +\ii A_1(x)\right)S(x)= 0,\quad S(0)=1.
\eqend
This allows us to write Gauss' law as
\eq
\tilde{G}(x) = -\partial_1\tilde{\Pi}_1(x) + \tilde{\rho}(x) \simeq 0
\eqend
with $\tilde{G}(x)\equiv\; \normal{S(x)^{-1}G(x)S(x)}$ and similarly
for $\Pi_1$ and $\rho$. In this form, Gauss' law is easily solved:
$\tilde{\Pi}_1(x)\simeq \Pi_1(0)+R(x)$ with
$R(x)=\int_{0}^x\dd{y}\tilde{\rho}(y)$. Introducing $p\equiv
\int_{0}^{2\pi} \dd{x}\tilde{\Pi}_1(x)/2\pi$ we obtain
\eq
\tilde{\Pi}_1(x) \simeq p + \bar{R}(x), \quad \bar{R}(x)=
R(x)-\int_{0}^{2\pi} \f{\dd{y}}{2\pi}R(x),
\eqend
and from $\Pi_1(0)=\Pi_1(2\pi)$ we get the constraint
\eq
\label{constr}
p + \bar{R}(2\pi)=\; \normal{q^{-1}\left( p + \bar{R}(0) \right) q }
\eqend
with $q\equiv S(2\pi)$ identical with the Wilson loop $W[A_1]$.  Due
to the cyclicity of the trace, we obtain from \Ref{a1} the effective
Hamiltonian
\eqa
\label{Heff1}
\HH'_{\rm eff}
=\int_0^{2\pi}\dd{x}\left(\tra{e^2\tilde{\Pi}_1(x)^2}+
\tilde{\psi}^*(x)\gamma_5
(-\ii\partial_1) \tilde{\psi}(x)\right) \nonu =
e^2 2\pi\tra{p^2} +
\int_0^{2\pi}\dd{x}\left(\tra{e^2\bar{R}(x)^2}+\tilde{\psi}^*(x)\gamma_5
(-\ii\partial_1) \tilde{\psi}(x)\right)
\eqaend
with $\tilde{\psi}(x)\equiv S(x)^{-1}\psi(x)$.  As $\tilde{\rho}^a(x)
= 2\tra{T^aS(x)^{-1}T^bS(x)}\psi^*(x)T^b\psi(x)$, it follows from the
invariance of $T^a\otimes T^a$,
\[
T^a_{AB}T^a_{CD}= (h^{-1}T^ah)_{AB}(h^{-1}T^ah)_{CD}\quad \forall
h\in\GG,
\]
that
\eq
\tilde{\rho}^a(x)  = \tilde{\psi}^*(x)T^a\tilde{\psi}(x).
\eqend
Note that the fermion fields $\tilde{\psi}^{(*)}$ still obey the canonical
anticommutator relations, but they have twisted boundary conditions
\eq
\label{bc}
\tilde{\psi}(2\pi)=q^{-1}\tilde{\psi}(0).
\eqend
It is easy to see that the Wilson loop variables obey
\eq
\label{loop}
\ccr{p^a}{q}=-qT^a,\quad \ccr{p^a}{q^{-1}}=T^aq^{-1}.
\eqend
The canonical anticommutator relations imply
$\ccr{\bar{R}^a(0)}{\tilde{\psi}^*(x)}=(\f{x}{2\pi}-1)\tilde{\psi}^*(x)T^a$,
and with $p^a=\Pi_1^a(0)-\bar{R}^a(0)$ we obtain
\eq
\label{WLF}
\ccr{p^a}{\tilde{\psi}^*(x)}=-\f{x}{2\pi}\tilde{\psi}^*(x)T^a.
\eqend
At this stage we have formulated the model entirely in terms of the
Wilson loop variables $p=p^aT^a$, $q$, and the twisted fermion fields
$\tilde{\psi}^{(*)}$ obeying the relations given above. $\HH'_{\rm
eff}$ has to be regarded as an operator on the tensor product
$\cH'_{\rm phys}$ of the Hilbert space of the Wilson loop variables
\cite{Dowker} and an appropriate fermion Fock space.

$\cH'_{\rm phys}$ is not quite the physical Hilbert space of the model
as we still have the constraint \Ref{constr} which is associated with
the invariance under rigid gauge transformations
\eqa
\label{rigid}
q&\to& h^{-1}qh\nonu
p&\to& h^{-1}ph\nonu
\tilde{\psi}(x)&\to& h^{-1}\tilde{\psi}(x) \quad \forall h\in \GG .
\eqaend
Indeed, it is easy to check that eqs.\ \Ref{constr}, \Ref{Heff1},
\Ref{bc}, \Ref{loop}, and \Ref{WLF} are invariant under
\Ref{rigid}.

To eliminate this constraint, it is convenient to introduce an
algebraic basis in the Lie algebra $\gg$ of $\GG$ as follows. Let
$e_{ij}$ be the $N\times N$ matrix with the elements
$(e_{ij})_{kl}=\delta_{ik}\delta_{jl}$. We define
$H_i=e_{i,i}-e_{i+1,i+1}$ for $i=1,\ldots, N-1$, $E_1^+=e_{1,2},
E_2^+=e_{1,3},
\ldots, E_{N-1}^+=e_{1,N}, E_{N}^+=e_{2,3}, \ldots,
E_{\f{1}{2}N(N-1)}^+=e_{N-1,N}$, and $E_j^-=(E_j^+)^*$.  These
matrices obey
\eqa
\label{basis}
\ccr{H_i}{H_j} &=& 0\nonu
\ccr{H_i}{E^{\pm}_j} &=& \pm a_{ij}E^{\pm}_j \qquad\forall i,j
\eqaend
with $a_{ij}$ the elements of $(N-1)\times \f{1}{2}N(N-1)$ matrix
given by $a_{ij}=\delta_{ik(j)}-\delta_{il(j)}-\delta_{i+1,k(j)} +
\delta_{i+1,l(j)}$ where $k(j)$, $l(j)$ are determined from
$E_j^+=e_{k(j),l(j)}$. Moreover,
\eqa
H_i^*=H_i,\quad &&(E_i^+)^*=E_i^- ,\nonu
\tra{H_iH_j}=b_{ij}&&\tra{E^+_iE^-_j}=\delta_{ij},\nonu
\tra{H_iE^\pm_j}&=&\tra{E^\pm_iE^\pm_j}=0\qquad \forall i,j
\eqaend
where $b$ is the $(N-1)\times(N-1)$ matrix with elements
$b_{ij}=2\delta_{ij}-\delta_{i+1,j}-\delta_{i,j+1}$. Note that $b$ is
invertible.\footnote{by induction one can show that its
determinant is $N\neq 0$} Obviously the matrices $H_i$, $i=1,\ldots,
N-1$ and $E_{j}^\pm$, $j=1,\ldots ,\f{1}{2}N(N-1)$, span the Lie
algebra of $\GG=\SU(N)$.  For $\GG=\U(N)$ we have to add to these
$H_0=1$ (the $N\times N$ unit matrix) and set $a_{0j}=0$,
$b_{0j}=b_{j0}=N\delta_{j0}$ so that
the relations above still hold. Then we can write
\[X=\sum_j\left(X^{0,j}H_j+ X^{+,j}E^-_j + X^{-,j}E^+_j\right)
\quad \forall X\in\gg \]
with $X^{0,j}= \sum_k(b^{-1})^{jk}\tra{H_kX}=(X^{0,j})^*$ with
$(b^{-1})^{jk}$ the elements of the inverse matrix of $b$, and
$X^{\pm,j}=\tra{E^\pm_j X} = (X^{\mp,j})^*$.

Every $q\in\GG$ can be represented as $q=h_q d h_q^{-1}$ with $d$ in
the Cartan subgroup of $\GG$, $d=\exp{(-2\pi\ii\sum_i Y^iH_i)}$ (note
that this representation is not unique).  We can explicitly solve the
constraint \Ref{constr} by performing a rigid gauge transformation
\Ref{rigid} with $h=h_q$. Denoting for simplicity the gauge
transformed $q$, $\tilde{\psi}$ etc.\ by the same symbol, we can
represent $\tilde{\psi}$ as
\eq
\tilde{\psi}(x)\simeq d^{-x/2\pi}\psi(x)
\eqend
with $\psi(x)$ free fermion fields obeying periodic boundary
conditions. Performing a Fourier transform,
$\psi(x)=\sum_{n\in\Z}\hat{\psi}(n)\exp{(\ii nx)}/\sqrt{2\pi}$, and
using
\nonueqa
\exp{(-\ii\mbox{$\sum_i$}\alpha^iH_i)}H_j
\exp{(\ii\mbox{$\sum_i$}\alpha^iH_i)} &=& H_j\\
\exp{(-\ii\mbox{$\sum_i$}\alpha^iH_i)}E^\pm_j
\exp{(\ii\mbox{$\sum_i$}\alpha^iH_i)} &=&
E^\pm_j\exp{(\mp\ii\mbox{$\sum_i$}\alpha^ia_{ij})}
\nonueqaend
for all real $\alpha^i$ following from \Ref{basis}, we obtain from
\Ref{constr} by a straightforward but tedious calculation that
$\hat{\rho}^{0,j}(0)\simeq 0$ and
\[
(1-\ee{\mp 2\pi\ii\mbox{$\sum_i$} Y^ia_{ij}})
\left( p^{\pm,j} + \f{1}{4\pi^2}\sum_{n\in\Z}
\f{\hat{\rho}^{\pm,j}(n)(\ee{\mp 2\pi\ii\mbox{$\sum_i$}Y^ia_{ij}}-1)
}{(n\mp\mbox{$\sum_i$}Y^ia_{ij})^2}\right) \simeq 0
\]
where $\hat{\rho}(n)=\int_0^{2\pi}\dd{x}\rho(x)\exp{(-\ii nx)}$, and
with that
\nonueqa
\tilde{\Pi}(x)\simeq p+\bar{R}(x)\simeq \sum_j p^{0,j}H_j +
\f{1}{2\pi}
\sum_{n\in\Z}\sum_j\left( \f{\hat{\rho}^{0,j}(n)}{\ii n}
\ee{\ii nx} H_j
(1-\delta_{n0}) \right.\\\left.  +
\f{\hat{\rho}^{+,j}(n)}{\ii(n-\mbox{$\sum_i$}Y^ia_{ij})}
\ee{\ii(n-\mbox{$\sum_i$}Y^ia_{ij})x}E^-_j +
\f{\hat{\rho}^{-,j}(n)}{\ii(n+\mbox{$\sum_i$}Y^ia_{ij})}
\ee{\ii(n+\mbox{$\sum_i$}Y^ia_{ij})x}E^+_j\right).
\nonueqaend
{}From \Ref{loop} we get
$\ccr{p^{0,j}}{d}=-d\f{1}{2}\sum_{k}(b^{-1})^{jk}H_k$.\footnote{we used
$T^a_{\AA\BB}T^a_{\CC\DD}= \f{1}{2} \delta_{\AA\DD}
\delta_{\BB\CC} - \alpha\delta_{\AA\BB}\delta_{\CC\DD}$ with
$\alpha=\f{1}{2N}$ for $\GG=\SU(N)$ and $0$ for $\GG=\U(N)$}
Hence we can represent the $p^{0,j}$ as $\f{1}{4\pi\ii}\sum_k
(b^{-1})^{jk}\f{\partial}{\partial Y^k}$. Putting this all together,
we obtain from \Ref{Heff1}
\eqa
\label{final}
\HH_{\rm eff}=-\f{e^2}{8\pi}
\sum_j\f{\partial^2}{\partial Y_j \partial Y^j} +
\sum_{n\in\Z}\hat{\psi}^*(n)
\gamma_5(n+\sum_jY^jH_j)\hat{\psi}(n) + \nonu
\f{e^2}{2\pi}\sum_{n\in\Z}\sum_j\left( (1-\delta_{0n})
\f{\hat{\rho}^{0,j}(n)\hat{\rho}^0_j(-n)}{n^2} +
\f{\hat{\rho}^{+,j}(n)\hat{\rho}^{-,j}(-n)}
{(n-\mbox{$\sum_i$}Y^ia_{ij})^2} +
\f{\hat{\rho}^{-,j}(n)\hat{\rho}^{+,j}(-n)}
{(n+\mbox{$\sum_i$}Y^ia_{ij})^2}
 \right)
\eqaend
with $\f{\partial}{\partial Y_j}\equiv
\sum_k(b^{-1})^{jk}\f{\partial}{\partial Y^k}$ and $\hat{\rho}^0_j\equiv
\sum_k b_{jk}\hat{\rho}^{0,k}$. This is our final result.

It should be pointed out that \Ref{final} is identical with the
Hamiltonian one obtains by fixing the gauge in eqs. \Ref{a1}, \Ref{a2}
to
\eq
\label{gauge}
A_0(x)=0,\quad A_1(x)= \sum_jY^jH_j
\eqend
which is essentially the Coulomb gauge. Thus the construction above
can be regarded as an justification of that gauge.

It is important to note that though we have eliminated in eq.\
\Ref{final} as many gauge degrees of freedom as possible, there are
still gauge transformations left, namely all those that leave
the diagonalized Wilson loop $d$ in the Cartan subgroup of $\GG$. For
example, $h(x)=\exp{(\ii x\mbox{$\sum_j$}\nu^jH_j)}$ with $\nu^j\in\Z$
is a gauge transformation (i.e. $h(0)=h(2\pi)=1$) leaving $d$
invariant but acting non-trivially on the $Y^j$: $Y^j\to Y^j+\nu^j$
(it is straightforward to check that \Ref{final} is invariant under
these gauge transformations). This can be understood as Gribov
ambiguity
\cite{G}: the Coulomb gauge condition $\partial A_1(x)/\partial x = 0$
does not uniquely determine one representative in each gauge orbit.
Demanding that the physical states of the model are invariant also
under these discrete gauge transformations leads to a highly
non-trivial vacuum structure on the full quantum level similar to the
one of the Schwinger model \cite{Manton}. We intent to report on that in
a future publication.

\end{document}